\documentclass[11pt]{article}
\usepackage{graphicx}
\usepackage{hyperref}

\makeatletter

\def\tableofcontents{\section*{\centering \contentsname
\@mkboth{\uppercase{\contentsname}}{\uppercase{\contentsname}}}
\@starttoc{toc}}

\renewcommand
    {\l@subsection}[2]{\addpenalty{\@secpenalty} \addvspace{0.5em plus 1pt}
\@tempdima 1.5em \begingroup
 \parindent \z@ \rightskip \@pnumwidth
 \parfillskip - \@pnumwidth
     \leavevmode \advance\leftskip\@tempdima #1 \hfil
\nobreak\hbox to\@pnumwidth{\hss #2}\par
 \endgroup}

 \makeatother

\begin{document}
\renewcommand{\abstractname}{}

\baselineskip=20pt         

\thispagestyle{plain}
\begin{center}
\LARGE \bf On the problem of the interpretation\\[4pt]
of quantum physics
\end{center}

\vspace{4pt}
\begin{center}
\large A.\,A. Grib\footnote{
{\bf A.A.\,Grib} \ Herzen Russian State Pedagogical University,\\
nab. r. Moiki 48, 191186  St.\,Petersburg, Russian Federation\\
E-mail:\, {Andrei\_Grib@mail.ru}}
\end{center}

\vspace{-17pt}
    \begin{abstract} {\noindent
\underline{\bf Abstract.} \,
{\bf Current thinking on the interpretation of quantum physics is reviewed, with
special detail given to the Copenhagen and Everett many-worlds interpretations.}
}\\[4pt]
{\bf {Methodological Notes.}}
\end{abstract}


{\small \baselineskip=7pt
\tableofcontents
}

\vspace{7pt}
\section{Introduction}
\label{Intr}

    In an earlier paper~\cite{1}, we made an attempt to show --- by
reviewing Russian physics textbooks and in the light of recent
theoretical developments (Bell inequalities, the Kochen--Specker
theorem~\cite{2}--\cite{6}) and the new Bell inequality violation
and teleportation experiments of Aspect, Zeilinger, and
others~\cite{7,8} --- how quantum physics should not be interpreted today,
the beginning of the 21st century. We refrained,
however, from discussing how it should be interpreted. It is
the purpose of this paper to address this issue, presenting
some constructive considerations about such an interpretation.

    Let us note, to begin with, that microworld phenomena,
from elementary particles to molecules, as well as the
phenomena of superconductivity and superfluidity, are
described so successfully by quantum theory that no one in
the physics community has any doubts as to the validity of its
mathematical formalism, which has demonstrated such
brilliant predictive power. It is attempts to interpret this
formalism, i.e., to explain it in terms of ordinary language,
which has split the physics community. The dissatisfaction
with the philosophical implications of this formalism has
prompted some physicists to try to change it in some ways: by
introducing hidden parameters, by renouncing the superposition principle,
or by extending quantum theory by
introducing decoherent histories~\cite{9}--\cite{12}. These alterations
have failed, however, to produce a theory that could compete
with that taught by textbooks, so interpretations that suggest
changing the mathematical formalism of quantum physics
will not be considered here. This leaves us with the following
three possibilities:

1. {\it Pragmatic interpretation.} The rules of quantum theory
allow highly accurate predictions to be made about the results
of our observations, and this is enough for us. We will simply
accept them as given and ask no questions as to why they are
as they are and not otherwise.

2. {\it Copenhagen interpretation.} This interpretation comes
in a number of versions, one of which is item 1. The present
paper discusses one relatively recent formulation of the
Copenhagen interpretation, put forth by Niels Bohr's son
Aage Bohr~\cite{13}, who summarizes many years of the interpretation debate
on quantum mechanics in Copenhagen in the latter half of the 20th century.

3. {\it Everett many-worlds interpretation.} First proposed in
1957 by H Everett~\cite{14} and propagated by J Wheeler~\cite{15},
B DeWitt~\cite{16} and others, the many-worlds interpretation is
used by such cosmologists as S Hawking~\cite{17, 18}, J Hartle,
and A Vilenkin~\cite{19} in their attempts to explain the quantum
birth of the Universe based on the idea of the Universe's wave
function existing in the absence of an observer. In the Russian
literature, UFN discussion papers by M B Menskii~\cite{20, 21}
should be mentioned.

\vspace{11pt}
\section{Basic postulates of quantum theory}
\label{Sec2}

Below, we list the main textbook postulates of quantum
theory. (Some revision of these postulates in the case of the
Everett interpretation leaves the mathematics unchanged.)

    (1) The state of a quantum object (for simplicity, the
object will be assumed to be a particle) is defined by its wave
function $\Psi (x,y,z,t)$, which is a function of Minkowski
spacetime coordinates and which is a Hilbert space vector
transformable as a unitary representation of the Poincar\'{e}
group under transformations from this group.

    (2) The observed properties of a quantum object are
described by self-adjoint operators in a Hilbert space. The
set of these operators includes, first of all, the following
Poincar\'{e} group generators: the Hamiltonian (translation in
time), momentum (translation in space), angular momentum
and spin (three spatial rotations), and three space coordinate
operators as a nonrelativistic limit of the three Lorentz
rotations. Importantly, as pointed out in review~\cite{13}, the
noncommutativity of --- and hence the Heisenberg uncertainty principle
for --- coordinates and momentum follows
from the noncommutativity of transformations of spatial
translations and Lorentz rotations, which points to a direct
relationship between quantum physics and the theory of
relativity. Without Minkowski spacetime, quantum physics
would be, generally speaking, entirely different from what it is
now.

    Besides the ones listed above, the set of observables
generally includes those without classical analogs: space and
time parity (space reflection and time reversal operations
from the full Poincar\'{e} group), charge parity, and various
kinds of charge (electric, baryon, lepton, strange, charm,
beauty, height, etc.). Finally, among the observables there
are local operators expressible in terms of local quantized
fields. For the electromagnetic field and other material fields,
the observables are the fields themselves, whereas for
electron--positron, baryon--antibaryon and other nonmaterial fields,
the observables are bilinear combinations of the field operators.

    I. In the absence of measurements, the wave function
varies with time in a deterministic way according to the
Schr\"{o}dinger equation
$$
i\hbar \frac{\partial \Psi }{\partial t} =\hat{H}{\kern 1pt} \Psi ,
$$
and the wave function is transformed through the unitary transformation
$$
\Psi (t_{0} )\to \Psi (t)=U(t-t_{0} )\Psi (t_{0} )=
\exp \left[ -i\hat{H}(t-t_{0} ) \right] \Psi (t_{0} ) .
$$

In the Heisenberg representation, the wave function does
not vary with time; what does vary is observable operators:
$\hat{A}\to U\hat{A}U^{-1} $.

    II. {\it Wave packet reduction postulate.}
    When measuring the observable $\hat{A}$ with an instrument,
the wave function $\Psi (t)$ collapses into one of the eigenfunctions
of operator $\hat{A}$, so that if
$$
\Psi =\sum \limits_n c_{n} u_{n}  ,
$$
where
$$
\hat{A}u_{n} =\lambda _{n} u_{n} ,
$$
then the probability that the instrument measuring property
$A$ will provide a reading $\lambda _{n} $ is given by the Born rule:
$$
w_{n} = \left|(u_{n} ,\Psi )\right|^2 .
$$
It is important to note that it is precisely the reduction
process~\cite{22, 23} which gives rise to randomness, i.e., indeterminism.
Into which function $u_{n} $ of the set $\{ u_n \}$ with
$|c_n |^2 \ne 0$ the wave function $\Psi (t)$ of a quantum object
transforms is totally
unpredictable. Here is von Neumann's well-known quotation on this point:
in quantum mechanics, ``dispersions are not due to our lack of knowledge
about the state, but to nature itself, which has disregarded the principle
of sufficient reason''~(\cite{24}, p. 302).
In quantum physics, unlike classical physics, probability is not a consequence
of our lack of knowledge, but rather it is objective probability that describes
objective randomness. The wave function `programs' the
behavior of a quantum object, preventing it from making a
transition to those $u_{n} $ for which $w_{n} =|c_{n} |^{2} =0$.
    In experimental terms, probability $|c_{n} |^2 $ is the frequency of
the event $\lambda _{n} $ (which is the eigenvalue of the operator $\hat{A}$
for the eigenfunction) observed over a large number of repetitions of
the experiment.
    Now what is this repetition?

    A particle, for example, an electron, is `prepared' in a
certain state $\Psi (x, y, z, t_0 )$., and then its property $A$ is
measured. Then we take the particle of the same kind in the
same state and measure $A$ again. The experiment is then
repeated a large (ideally, an infinite) number of times. As a
result, according to von Mises's statistical definition of
probability, the experimentalist will learn the frequencies
with which the $\lambda_1 ,\lambda_2 ,{\rm ...}$ are observed,
applying the Born rule.

    Thus, confirming predictions extracted from the wave function generally
requires considering an ensemble of identically prepared particles.

    III. {\it Superposition principle.}
    If $\Psi_1 $ and $\Psi_2 $ are the states of
the system, then any linear combination thereof,
$\Psi =c_1 \Psi_1 + c_2 \Psi_2 $, with $c_1$, $c_2$ being complex numbers,
is also a possible state of the system.

    IV. {\it Corpuscular-wave dualism.}
    The wave function of a quantum particle can be obtained by applying
to a Fock vacuum state the operator of the local quantum field
associated with the particle (the electron--positron field for
the electron and the positron, the electromagnetic field for the
photon, etc.). It is then clear why the wave function satisfies
the Schr\"{o}dinger equation and what the reason for and
meaning of corpuscular-wave dualism are. The reason is
that, even though we observe the particle as a pointlike
object, the program for its evolution in time is determined
by a quantized field that exists, in reality, as an element of the
set of observables of a quantum object. Multiplying the
absolute square of the wave function, which is the spatial
probability distribution, by the electron charge yields (to
within a normalization factor) the charge density of the
electron field. Observing this density requires a state of an
infinite number of electrons (which is an eigenstate for the
frequency operator; see below), so that the wave properties of
the electron are also observed as the properties of this
ensemble of particles. In the truth, as shown by Everett~\cite{14},
Graham~\cite{25}, and Hartle~\cite{26}, Born's rule allows a proof.

Indeed, consider an infinite set of states of a particle that
have been identically prepared at different times (indexed 1, 2, \ldots)
and also consider a wave function which is a von Neumann's
infinite-dimensional tensor product:
\begin{equation} \label{EQ1}
|\left. \Psi ^{\infty } \right\rangle =
|\left. \Psi {\rm ,1}\right\rangle \otimes |\left.
\Psi {\rm ,2}\right\rangle \otimes \ldots .
\end{equation}
    It can then be shown~\cite{25, 26} that the vector
${\left| \Psi ^{\infty }  \right\rangle} $ is an
eigenfunction of a special operator $\hat{f}^{(k)} $, whose eigenvalue is
the frequency of the outcome $\lambda _{k} $ obtained in measuring the
property $A$ ($\hat{A}u_{n} =\lambda _{n} u_{n} $), so that
    \begin{equation} \label{EQ2}
\hat{f}^{(k)} \left| \Psi ^{\infty }  \right\rangle =
\left| \left\langle k | \Psi  \right\rangle \right|^2
\left| \Psi ^{\infty }  \right\rangle .
\end{equation}
    Here, we have resorted to the Dirac notation
$ |\left\langle \left. k|\Psi \right\rangle \right. |^2 =|(u_{k} ,\Psi )|^2 $.

    Given the above discussion of quantum physics postulates, what is
the problem that splits the physics community?
    The answer is measurement. There are two key features to this concept.

    A. {\it Conversion of a pure state into a mixture of states.}
    The question arises: Why is it that a quantum object, when not
measured, evolves in such a way that its wave function
changes deterministically in accordance with the Schr\"{o}dinger equation
--- a process which is described by a unitary
operator $U(t)$ --- whereas in making a measurement the wave
function collapses unpredictably into an eigenfunction of the
observed quantity? Is it not true that the measurement, i.e.,
the interaction of the probe particle with the apparatus which,
in turn, also consists of particles, is also described by the
Schr\"{o}dinger equation, i.e., by a unitary operator?

    The following example will serve to illustrate why a
measurement (or rather a first-kind measurement) violates
unitary of the evolution.

    Suppose there is a particle with a wave function
$\Psi (x,y,z,t) $ and an instrument
measuring the observable $A$, resided initially in the state
$\chi _{0} (x_{a} ,y_{a} ,z_{a} ,t)$.
    Upon making a measurement, considering the particle and the instrument to
be quantum in nature, we should have
\begin{eqnarray} \nonumber
\Psi ( x,y,z,x_{a} ,y_{a} ,z_{a} ,t ) &=&
\left( \sum \limits_{n} c_{n} u_{n} (x, y ,z, t ) \right)
\chi_0 (x_a ,y_a ,z_a ,t )
\\    \label{EQ3}
&\to& \sum \limits_{n} c_n \, u_n (x,y,z,t') \, \chi_n (x_{a} ,y_{a} ,z_{a} ,t') ,
\end{eqnarray}
    where the functions $\chi_n $ correspond to various positions of the
instrument `pointer'.

    Transition~(\ref{EQ3}) requires a special kind of interaction, because we
would generally have an expansion in terms of different basis functions:
    \begin{equation} \label{EQ4}
\sum \limits_{n,k} c_{n k} u_n \chi_k  ,
\end{equation}
    which does not lead to a measurement.

    Thus, transition~(\ref{EQ3}) proceeds unitary and occurs according to
the Schr\"{o}dinger equation.

    The question to address next is: If the composite particle--instrument
system is described at the time moment $t$ by the function~(\ref{EQ3}), what
is it that describes the particle as a subsystem? The answer is that, for
orthogonal $ u_{n} $, the subsystem has the density matrix
$$
\rho = {\rm diag} \left( |c_1 |^2 , |c_2 |^2 , |c_3 |^2, \ldots \right),
$$
    which is diagonal in the given basis.
    Now, what is the interpretation of this density matrix?

    For an observer watching the subsystem, the density
matrix $\rho $ describes a mixture of states, i.e., the observer will
say that the particle is in a certain one of the set of states
$u_1 ,u_2 , \ldots , u_N $, the respective probabilities of the outcomes
$u_1, u_2, \ldots $ being $|c_1|^2, |c_2|^2, \ldots $.
    However, in this interpretation the system as a whole is no longer
described by a pure state, superposition~(\ref{EQ3}), but only by a mixture
of states, the particle--instrument system being in states
$u_1, \chi_1, \ u_2, \chi_2, \ldots $ with respective probabilities
$|c_1|^2, |c_2|^2, \ldots $.

    Thus, everything depends on whether or not the observer
looks at the subsystem. If not --- i.e., if only the system as a
whole is observed --- the states do not mix, and the system is,
as before, described by a pure state.

    Note that the very possibility of observation is important
here. It is relative to the observer that a pure state turns into a
mixed state, a situation which is, as de Broglie put it, the
complimentarity of whole and part: knowing a part destroys the whole.

    The conversion of a pure state described by a single
wave function into a mixture of states with different wave
functions constitutes a nonunitary operation contradictory
to the Schr\"{o}dinger equation. The net result of measurement
is that the observer now knows (`registers') in which one of
all the possible states the system ends up, i.e., of all the
different $u_n \chi_n $ states only one --- that which has been
actually realized --- is fixed. This last registration event is
completely analogous to that in the classical physics, in
which, after a measurement has been performed, the
observer knows for certain that the instrument pointer has
a definite position even though they do not know which
one. But after the observer has looked at the instrument,
only one of all the possibilities occurs. What makes things
mysterious is the conversion of a pure state into a mixture,
something which violates the ordinary unitary evolution of
a quantum system.

    In his book~\cite{27}, V A Fock argues, quite rightly, that these
two types of change undergone by the wave function reflect
two types of change also occurring in classical physics: a
change according to the equation of motion, and a change in
the initial conditions.

    The reduction of the wave packet leads, through the
interference of the measuring instrument, to a change in the
initial conditions for the subsequent evolution. Book~\cite{27},
however, makes no mention of the fact that in classical
physics this change in the initial conditions can be described
in terms of the same classical physics. In quantum physics,
however, this is not the case: the result of measurement
cannot be obtained from the Schr\"{o}dinger equation.
The above problem was a subject of discussion.

    The above problem was a subject of discussion between
the present author and Vladimir Aleksandrovich Fock for
two years after the publication of his book~\cite{27}, two years
which, alas, proved to be V A's last. There were plans to
compile the Russian translations of major works on measurement
problems in one book some day, with a preface on V A's own position on the
subject. Unfortunately, these plans were cut short by his death.

    B.\, Another problem concerning measurement is related
to its special aspect in that it {\it converts the numerically indefinite
values of a physical quantity into a definite one.}
    Unlike classical physics, this conversion is not simply the fixation of
the value that existed prior to the observation and was only unknown
to the observer, but rather the conversion of an `objective
indeterminacy' to a numerical value.

    As we noted in Refs~\cite{28}--\cite{30}, the experimentally con-
firmed violation of Bell's inequalities in quantum physics
forbids speaking of the existence, prior to measurement, of
the numerically definite values of properties described by
noncommuting operators. Measurement does not fix the
existence of a value that existed prior to it; it creates this value.

    Another example to illustrate the nonexistence, prior to
measurement, of the numerical values of observables (for a
one- or more dimensional Hilbert space) is the Kochen--Specker example
--- a so-called quantum polyhedron~\cite{3}.

    We will now proceed by considering those interpretations
of quantum physics which, in some way or another, accept its
mathematical formalism. The pragmatic interpretation,
which we already mentioned in the Introduction, is essentially that
we should consider the quantum postulates as just
the way things are and stop at this. But if we do not want to,
there are two possibilities to consider, which are outlined in
Sections \hyperref[Sec3]{3}  and \hyperref[Sec4]{4}.

\vspace{11pt}
\section{Copenhagen interpretation}
\label{Sec3}

    Originally proposed by Niels Bohr, Werner Heisenberg, and John von
Neumann and further developed by Aage Bohr in the latter half of the
20th century~\cite{13}, the Copenhagen interpretation can be summarized
as the following set of principles.

    The world is divided into quantum objects and apparatuses, the former
being described by quantum physics and the latter by classical physics
[so that the results of measurements are described in terms of classical
concepts and classical (Boolean) logic].
    Notice that, while most of apparatuses themselves are macroobjects
consisting of a large number of microparticles, this is not necessarily
the case.
    In the Stern--Gerlach experiment, which measures the projection of
the electron spin onto a certain direction, the role of the apparatus is played
by silver ions which move in this direction or another along a
classical trajectory, depending on the direction of the magnetic
field and the value of the spin projection.
    While silver ions are not macroscopic bodies, under the conditions of
the experiment their center of inertia moves along a classical trajectory and
plays the role of a classical apparatus.
    The main property of the apparatus is that it enables the observer to
measure only commuting observables, those which provide him with information
about some property of the quantum object even though the apparatus consists
of quantum objects.
    But of course, ultimately, even the quasiclassical ion trajectory is
registered by a macroscopic apparatus --- by a photographic film, for example.

    The usual choice for such commuting apparatus observables is
quasiclassical macro-observables, so that, due to
decoherence phenomenon (which depends on the number of
particles in the environment) the density matrix of the
quantum object rapidly becomes diagonal. Those properties
of quantum objects that are described by noncommuting
operators take on their numerical values in additional
experiments run with various instruments, but, if not via
measurement, there are no fixed numbers to characterize the
properties of a quantum object. Bohr's and Fock's concept of
relativity with respect to the means of observation, which is
about the `emergence' of numerically definite properties upon
making a measurement, is the most relevant to describe this
situation. This concept first appeared in the special theory of
relativity (STR) where, unlike Newtonian mechanics, the
length of an object and the duration of a measurement
process are not the attributes of the object itself but rather
refer to the `relation' of the observed object to another object
associated with the observer --- to an inertial reference frame.
    A reference frame, as a collection of rulers and clocks, is an
`apparatus' that measures length and duration of the process
as the `relation' to this apparatus. These `relations' change as
the reference frame changes, which manifests itself in the
Lorentz contraction of the length scale and time. There is, of
course, a `proper' reference frame, which reflects the `self-relation';
nonetheless, already in STR one cannot speak of a definite length and
a definite duration without identifying the reference frame.

    In quantum physics, a quantum object is `in and of itself'
described by operators rather than by numbers (as in classical
physics), so that it `objectively' represents a set of observable
operators. As mentioned in Section \hyperref[Sec2]{2}, these operators
include generators of the Poincar\'{e} group, those of local quantized
fields and various gauge transformation charges.

    With respect to an apparatus described classically --- that
is, in the language of numbers --- operators `turn into'
numbers, i.e., into the eigenvalues of the operators. The
wave function characterizes a certain special `relation'
between an apparatus and the quantum object and, in this
sense, describes both of them together. The wave function is
defined in combination with the `preparing' or `measuring'
apparatus. Reference~\cite{31} is another work to highlight the
importance in modern physics of such a category as `relation'.
The situation that arose in quantum physics is occasionally
described by some authors~\cite{32} in terms of there being
`objective indeterminacy', `objective randomness', and
`objective probability' in the microworld. To reemphasize, a
quantum object `in itself' is not generally characterized by any
number. For example, the presence of a coordinate operator
instead of a number indicating the position of the object can
be interpreted as `objective indeterminacy' of the position of
the quantum object in space. It is only through measurement
that such a number-defined position emerges. Unlike
`subjective indeterminacy', the term `objective indeterminacy' implies that
indeterminacy is not related to our lack of
knowledge (as is the case in classical physics) but rather is a
property of the object itself. Heisenberg's words~\cite{33}, speaking of
quantum reality as ``something in itself'' brings quantum physics closer
to Platonism.

    Description on the language of operators implies, in a
sense, the existence of `a coordinate in general', of `a
momentum in general', of `a particle in general' (the principle
of identity of particles), etc. This is, in fact, medieval realism,
known for its opposition to nominalism on the question of the
existence of general notions (say, a man in general or a woman
in general --- not only specific Bobs or Alices). As is known, it
was nominalism which got the upper hand.

    But in quantum mechanics, the principle of identity of
particles states that the members of a system of electrons
cannot, due to their being identical, be assigned an individual
`name' such as first, second, etc., and the only thing we can
speak of is the number of electrons. All this calls for more
attention to realism and its ensuing Platonism.

    Finally, there is, according to Finkelstein~\cite{34}, yet another
language to speak. A quantum object is a set of plurality of
`acts' that are described by operators, but the apparatus turns
`acts' into `facts' that are characterized by commuting
operators and are therefore described by numbers.

    The time evolution of a quantum object is described in a
natural way in the Heisenberg representation. Thus, in the
quantum theory of fields, when describing the double
commutator of the Hamiltonian with the field, the time
evolution of the field leads in a natural way to quantized
field equations: the Klein--Fock equation for a scalar field, the
Dirac equation for a spinor field, etc. However, in the
Schr\"{o}dinger representation, which is usually equivalent to
the Heisenberg representation, one should speak of the
apparatus, without which the wave function is not defined.
From the quantum field theoretical perspective, the wave
function results from the application of the field operator to a
vacuum, and if the apparatus moves noninertially --- for
example, with a constant acceleration --- the Fulling--Unruh
effect arises~\cite{35} and equivalence is violated. A similar
nonequivalence also arises when as the Fock vacuum for the
operators of a field with a certain mass is taken the vacuum of
a field with a different mass~\cite{36}. This seems to make the
Heisenberg representation preferable for describing the
evolution of a quantum system.

    Now let us look in more detail at the problem of wave packet reduction.

    A number of attempts have been made to interpret this
concept. As is known, von Neumann~\cite{24} linked the reduction
phenomenon to the consciousness of the observer in the sense
that any apparatus extends the sense organs of the observer
who receives information about the world around them. Any
apparatus can be considered to consist of quantum objects ---
atoms and molecules --- and, hence, can be treated by the laws
of quantum physics --- assuming that the `principle of
transferring the boundary' is valid.

    The object--apparatus boundary can be brought toward
the observer's eyes, which are considered classical, and at the
same time, what was previously considered to be an apparatus
is now combined with the object being studied into a common
quantum system. According to von Neumann, the result of
the observation is the same wherever the boundary is drawn.
Although the boundary can be moved further into the brain
of the observer, consciousness or the `abstract I' will always
be present as a subject of cognition receiving information.

    The apparatus--object boundary may not necessarily be
drawn in the present. Because the apparatus is treated
classically, it is possible, by applying the laws of classical
physics with its determinism, to argue that the apparatus at
the time of measurement had produced --- even before the
observer looked at it --- a certain definite result, though
unknown to the observer. The logic here is that the observers
would see a quite definite outcome if they looked in the past
(not now!) at the object which they call the apparatus and
which interacted via a special kind of interaction (the
`measurement Hamiltonian') with the microobject.

    Von Neumann's view on the role of the subject of
cognition in the process of wave packet reduction has been
criticized for `solipsism'. Reference~\cite{37} argues, not without
irony, that, according to von Neumann's interpretation,
``every one of us should be of the view that he alone is the
observer of significance, whereas the rest of the Universe and
its residents satisfy the Schr\"{o}dinger equation at any time ---
except when he does the measurement.''

    There is a well-known philosophical argument~\cite{38} to
advance against this, though, according to which the notion
of a number does not apply to the subject of cognition. The
only subject of knowledge is I. All the rest is objects for the I.
It goes without saying that this I should not be identified with
me as Alice, Bob or some other objects both for myself and for
others. Megalomania occurs only when I, Alice, am the
Subject --- unlike Bob who is no more than an object. Every
cognizing observer is I as the subject of cognition. Because I is
the one and only, it follows that what I see as a result of
quantum observation will be the same for other observers.
This is why there is no place for the Wigner's-friend type
paradox, according to which, if the reduction event is due to
the consciousness of the observer, then different people ---
say, Wigner and his friend --- could see different results. This
paradox assumes a multitude of consciousnesses, which is
inconsistent with von Neumann's reasoning. Given his
awareness of and connection with the great German philosophical tradition,
it seems von Neumann knew well what he meant when using the term
`abstract I of the observer'.

    It can be said without any philosophical argumentation,
though, that, according to the rules of quantum mechanics,
for the first-kind observations --- in which the object after the
measurement resides in an eigenstate of the observable
operator --- there is no inconsistency between the measurements of two
observers if the measurements are conducted at different instants of time.
    Whether the measurement was first
done by Wigner and then by his friend, or the other way
round, the result would be the same for both.

    This view on the role of consciousness is very much shared
by Schr\"{o}dinger. To quote from his What is Life?~\cite{39}:
```I', when taken in the broadest sense of the word --- i.e., every
conscientious mind that sometimes said or felt `I' --- is nothing
other than a subject capable of controling the `motion of
atoms' according to the laws of nature.''

    Later on, F London and E Bauer~\cite{40}, E Wigner~\cite{41}, and
R Penrose~\cite{42} furthered the idea that it is the observer's
consciousness that causes the reduction of the wave packet.

    Here is the reasoning of London and Bauer. Suppose a
complex system --- quantum object X, apparatus Y, and
observer Z --- is described, following the interaction of the
quantum object with the apparatus, by the wave function
    \begin{equation} \label{EQ5}
{\left| \Psi (x,y,z) \right\rangle} = \sum _{k} a_{k}
\left|u_{k} (x)\right\rangle \otimes \left|v_{k} (y)\right\rangle
\otimes \left| w_{k} (z)\right\rangle ,
\end{equation}
    where $\left| w_{k} (z)\right\rangle $
are various states of the observer. It should be
noted that, unlike von Neumann's observer, the observer Z is
objectivized and described by a wave function. (Clearly, for
philosophers like Schopenhauer and N A Berdyaev, the
immediate reaction would be that what we are speaking of is
not the subject of cognition but consciousness as an inalienable property
of the observer.)

    The observer as a subsystem of the complex system is
described by the density matrix which, however, is not as
yet interpreted as a mixture of states. But here is what the
authors of Ref.~\cite{40} go on to say: ``For him (the observer),
only the object X and the apparatus Y are things from the
outside world. To himself, on the contrary, he is in a
special kind of relation: he has a well-known ability, which
can be called the ability of self-observation (introspection),
and so needs no mediator to identify the state he is in.
This self-cognition enables him to break the chain of
statistical links which are expressed by the wave function
$\sum _{k} a_{k}  \left| u_{k} (x) \right\rangle \otimes
\left| v_{k} (y)\right\rangle \otimes \left| w_{k} (z) \right\rangle $
and to say: I am in the state $\left| w_{k} (z) \right\rangle $,
or I see $g=g_{k} $\, ($g$~being the reading of the apparatus)
or directly $f=f_{k} $\, ($f$~being a property of the quantum object)''.

    Thus, it is the observer's consciousness which is responsible for
the transformation of the density matrix into a mixture of pure states,
so that introspection produces a new wave function for the object,
namely $ \left| u_{k}(x) \right\rangle $.
    Consciousness, unlike unconsciousness, can manifest itself in a certain
pure state from the set $w_{k} (z)$, but not in a state described by the
density matrix.

    It is relevant here to discuss the Wigner's friend paradox.
A transition into a pure state is not due to an objective,
outside-of-the-observer process, but due to something like the
observer's choice. But then, why cannot another observer
make a different choice? It is important to note, however, that
much depends on who is the first and who is the second when
do taking a measurement. If the reduction to a certain wave
function was performed by the first observer, then, according
to the properties of the measurement Hamiltonian, the second
observer will obtain the same wave function as the first when
taking the same measurement. (Everett~\cite{14} followed the same
reasoning when arguing why different observers find themselves in
one and the same of all possible Everett's worlds.)
Squires~\cite{43} and Menskii~\cite{21, 44, 45} closely follow this view
on the role of consciousness.

    Clearly, a proof of the validity of this interpretation
should be looked for in psychology and physiology, disciplines for which
the main problem yet to be solved is the
psychophysical one of how consciousness (that is, something
nonmaterial and described in psychological terms) controls
the body which is described in physical terms. Whether
quantum theory can help in solving this problem remains an
open question.

    An objection to this view may be seen in the question
R Penrose~\cite{46} asked regarding the Everett interpretation:
``How can we know that consciousness cannot be conscious
of itself as something described by the density matrix or by a
number of wave functions?''

    To add to the above, the indeterminism and probability
that arise during measurement are due to the observer; they
do not exist in the objective quantum world. But if this
probability is objective, then it reflects the uncontrollable
randomness in how the observer makes their choice. The
observer cannot control this randomness and, even though
the randomness has no relation to their lack of knowledge, as
in the classical physics, it is their choice that determines it.

    In this connection, the question posed by M B Menskii~\cite{21}
is worth considering: Is it possible that the observer, with
his subconscious access to all the possibilities offered by
superposition, can make a choice which benefits him?

    After repeating this many times, he will end up violating
the probabilistic predictions of quantum physics. While this
choice is possible in single individual cases, in the multiple
repetition scenario the action of consciousness cannot be
distinguished from a pure coincidence.

    Finally, if the observer is still unable to make such a
choice, it seems natural to conjecture that this choice will be
made by the microparticle --- but only with respect to the
observer who receives information about this particle. This is
reminiscent of the free will of the electron Dirac spoke of in
the early days of elaborating quantum physics.

    The role of reception of information by the observer is
particularly manifested in what is known as `negative'
experiments. The first such example was given by Renninger~\cite{47}.

    Suppose a source of charged particles (electrons) is placed
inside a sphere with a hole in it, coated with a scintillating
material, and surrounded by a larger sphere. An electron
striking the smaller sphere causes a light flash (scintillation),
and information about the occurrence (or otherwise) of the
flash on this sphere is transferred to the observer (by means of
electromagnetic radiation). Suppose, however, that although
an electron has flown out of the source, no scintillation was
observed. The observer will then say that the electron has
escaped through the hole. Due to the Heisenberg uncertainty
relation, the momentum of the escaping electron is different
from that in the case of no negative observation and also from
the momentum the electron had when leaving the source.
Thus, the reduction of the wave packet occurred even though
the electron was not observed in a certain region of space. A
similar macroscopic situation would arise if the observer
knew that a train was going to leave Chicago with a known
speed, maybe heading for Los Angeles, maybe heading for
New York, and a failure to observe the train passing an
intermediate station would change the speed of a train going
from New York.

    Unlike the defenders of the reduction model discussed
above, Roger Penrose advocates the idea of `objective
reduction' as a certain physical process violating the unitarity of
evolution~\cite{42, 46}. Given the notion of applying
relativity principle to the means of measurement, this is
reminiscent of the Lorentz--Fitzgerald idea that the Lorentz
contraction due to changing a reference frame can be
attributed to the contracting effect of electromagnetic forces
rather than being interpreted as a geometric property of the
Minkowski four-dimensional spacetime.

    Penrose suggests a process in quantum gravity to illustrate
his point. He does not believe that different spacetimes can
form a superposition and argues, therefore, that quantum
gravitation violates unitarity.

    Quantum gravitation should perhaps be taken into
account when considering interaction with macroscopic
bodies. That macroscopic bodies may require the quantization of
gravitation for their description is a long-discussed
idea. To illustrate, if we write out the de Broglie wavelength of
a truck (or rather of its center of mass), we find that the period
of such vibrations is much less than Planck's time, suggesting
the necessity of the quantization of gravitation.

    Unfortunately, quantum gravitation has not yet been
developed, so Penrose's idea remains just a hypothesis.

    Some support for the idea of considering consciousness
was drawn from a field which came to be known as `quantum
logic' (a subfield of quantum axiomatics).

    J von Neumann, the cofounder of quantum mechanics,
came across the question of why the properties of quantum
objects are described by Hilbert space operators~\cite{24}, Ch.~6.
    In his joint work~\cite{48} with Birkhoff, it has been found that this is
due to the existence of a special mathematical structure, a
complete orthomodular lattice (for an exact definition, see,
for example, book~\cite{28}), which possesses the properties of a
logical structure with operations `AND' (conjunction, $ \wedge $),
`OR' (disjunction, $\vee$), and `NOT' (negation, $\ne$), for which the
properties `AND', `OR' does not hold.
    In Aristotelian (or Boolean) logic, we always have
    \begin{equation} \label{EQ6}
A\wedge (B\vee C)=(A\wedge B)\vee (A\wedge C) .
\end{equation}
    However, if
    \begin{equation} \label{EQ7}
A\wedge (B\vee C)\ne (A\wedge B)\vee (A\wedge C) ,
\end{equation}
    this leads to quantum theory with its noncommuting operators.
    The original analysis in paper~\cite{48} treated the
simple case of finite-dimensional Hilbert space. Subsequent
work by Jauch~\cite{49}, Piron~\cite{50}, and others from the Swiss
theoretical school extended the analysis to include an infinite--dimensional
Hilbert space.

    Whereas the logic of our consciousness is distributive (or
Boolean), that operating in the world is of a different ---
nonhuman, non-Boolean --- nature. Logic, like geometry
with its many non-Euclidean versions, is not unique. Other
logics, like logics of the behavior of some objects, can be
revealed experimentally, for which purpose it is necessary to
design experiments in which the properties of the objects
could be measured in such a way that, at the same time, more
complex properties, considered as conjunction and disjunction of
the original properties, can also be established.
    Thereafter, it is also necessary to check whether the
distributivity property holds for conjunction and disjunction.

    Non-Boolean logic allows one to argue that a pointlike
electron with a definite momentum (property A) flying
through a screen with two slits (properties B, C), i.e.,
$A\wedge (B\vee C)$ is not the same as when an electron (while
remaining pointlike) flies either through slit C or through slit B.

    The following two aspects reveal the role of consciousness
in `quantum logic':

    (a) Boolean consciousness, receiving information about
the microworld, projects the non-Boolean quantum-logical
world onto a Boolean structure. The nonisomorphic nature of
these structures manifests itself in that at different instants of
time consciousness chooses some Boolean substructures
corresponding to the sets of the commuting operators of the
quantum object, so that the noncommuting operators are
measured at different instants of time in order to obtain
information about the quantum object. Note that time plays a
very important role here --- as if the observer would invented
time expressly to identify the quantum object~\cite{51};

    (b) Boolean consciousness defines the function of truth
(difference between true and false events) on the observables
of a Boolean subsystem. However, unlike the truth function
in classical physics, prior to observation this function is not
generally defined on a nondistributive lattice (Kochen--Specker
theorem~\cite{2}) and occurs randomly, which is the
reason for quantum indeterminacy and for relativity with
respect to observation.

    Now the question can be asked: If non-Boolean logic is
nonhuman, how can we think and speculate about it at all?

    Here we face a situation quite similar to that with extra
dimensions. Although the fourth and fifth dimensions are
beyond our imagination, analytical geometry makes it
possible to translate the question of multidimensional images
into that of algebraic equations that represent these images
without the need for geometrical imagination. It is appropriate here
to recall L D Landau's words that humans can
comprehend things which they cannot imagine. By the same
token, non-Boolean logic is studied by replacing logical
operations by lattice operations isomorphic to --- but not
identical to --- them.

\begin{figure}[ht]
\centering
   \includegraphics[width=45mm]{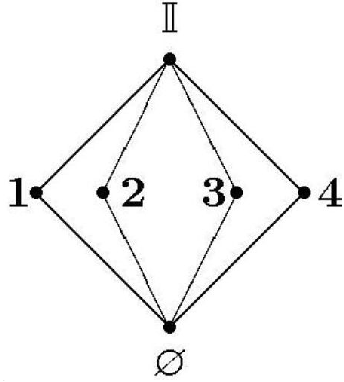}\\
\caption{Hasse diagram. Slashed circle \O \, denotes `always false', and I
denotes `always true'. Dots 1, 2, 3, 4, called logic atoms, denote:
1\,---\,$S_x = 1/2$, 2\,---\,$S_x = -1/2$, 3\,---\,$ S_y = 1/2$, and
4\,---\,$S_y = -1/2$.
    Lines converging at the bottom and at the top denote conjunction and
disjunction, respectively. Upward line stands for `it follows'.}
  \label{image}
\end{figure}

    A simple example of a particle with spin $S = 1/2$,
characterized by two spin projections onto the $x$-, $y$-axes~\cite{28},
will illustrate the above point without using complex
mathematical reasoning (which, in turn, assume knowledge
of the relevant definitions). Let us draw the Hasse diagram for
a quantum-logical lattice (Fig.~\ref{image}).
    From this diagram it is seen that
    \begin{equation} \label{EQ8}
1 \wedge 2 = 2 \wedge 3 = 3 \wedge 4 = 2 \wedge 4 = \O ,
\end{equation}
    \begin{equation} \label{EQ9}
1 \vee 2 = 2 \vee 3 = 3 \vee 4 = 1 \vee 4 = {\rm I} .
\end{equation}
    The distributivity property, however, does not hold:
    \begin{equation} \label{EQ10}
1 \wedge (2 \vee 3) = 1 \wedge {\rm I} = 1 \ne (1 \wedge 2) \vee
(1 \wedge 3) = \O \vee \O = \O .
\end{equation}

    (It is assumed, in accordance with the properties of the lattice,
that $1 \wedge I = 1$.) As a possible truth function for this simple
lattice, the observer could introduce, for example, the
following function: 1 is true and 2, 3, 4 are false as
incompatible with 1. Notice, incidentally, that for more
complex quantum systems, in particular, for those with spin
1, it may happen that, unlike a Boolean lattice, such a
function cannot be introduced at all (as Kochen--Specker
and Peres examples illustrate~\cite{3}). The simple example we
have given shows, however, that the falsity of 3 and 4 implies
that 3 or 4 are always true, which is, of course, at odds with
our logic.

    Therefore, a `Boolean observer' sees a Boolean sublattice
of elements, which in no way contradicts her/his logic. At the
next instant of time, she singles out another Boolean system,
3 and 4, and assigns a new truth function to it: 3 is true, 4 is
false or 4 is true, 3 is false. All this cannot be otherwise, given
her Boolean consciousness. She cannot, after all, allow a
situation with 3, 4 both false and 3 or 4 true. But, then, the
`Booleanization' of non-Boolean logic leads to an indeterminate definition
of truth and to relativity with respect to the means of measurement.

   Thus, it is untenable to argue that truth and falsity exist in
a quantum system prior to measurement and that all the
difference between a classical and a quantum object is
nondistributivity with observer-independent truth values.
One cannot speak of any definite measurement-independent
values of physical quantities described by noncommuting
operators.

    Nondistributivity, as is easily seen and as was realized
already by Birkhoff and von Neumann, is inconsistent with
Kolmogorov's probability axiomatics and requires introducing
the probability amplitude as a new tool for describing
randomness for non-Boolean lattices. It is beyond the scope
of this paper, though, to discuss this in more detail.

    The brief discussion above concerned the quantum-
logical version of the Copenhagen interpretation, a version
which enables one to say, in addition to what London and
Bauer, and von Neumann said, that the important function of
consciousness --- and one without which it would hardly
deserve its name --- is to distinguish between true and false.
But, true and false in Boolean classical physics are distinguished
`objectively' without any mention of consciousness.
In quantum physics, on the contrary, true and false emerge in
microobjects when their properties become realizable and do
not exist otherwise. However, what consciousness conceives
as true and false is determined not by consciousness itself but
by an object external to it, and is determined in a random way,
with a probability calculated on a Boolean substructure with
the aid of the wave function.

    There is, finally, a view, often associated with some
statements by Niels Bohr, that there is something in
macroscopic bodies which prevents applying quantum
theory to them to its full extent, so that consciousness plays no
special role and the readings of macroscopic instruments are
the same as they would be in the absence of it.

    At one time there was a belief that macroscopic bodies
differ from microscopic objects in that the former consist of a
large number of microscopic objects and that perhaps we
have a situation analogous to that of passage from statistical
physics to thermodynamics. Then classical physics, like
thermodynamics, should be more accurate as the number of
particles in the macroscopic object increases. This hope,
however, was dashed by Hepp's theorem~\cite{52} asserting that,
for any quantum system of many (but a finite number of)
particles, there is always an observable whose operator does
not commute with the rest of the operators, and whose
measurement will lead to an arbitrarily large departure from
classical theory. Only in the case of an actually infinite
number of particles is it possible to obtain a classical theory
in which interference effects are absent due to the emergence
of superselection rules. Real macroscopic objects always
consist of a finite number of microparticles.

    Another attempt to turn from considering microobjects to
considering macroobjects is based on the idea of decoherence.
The crux of the idea is essentially that a macrobody not only
consists of a large number of particles but it also interacts with
a large number of particles in its environment. If we take an
average over the ensemble of particles in the environment,
and if, in doing so, we focus our attention only on describing
the macrobody, then a certain system of mutually commuting
quasiclassical observables can be singled out so that the
evolution of such a system can be described on this basis to
good accuracy in terms of classical physics. So, if we are
interested to know the position of the pointer of the
macroapparatus, a distinction should be made between the
external environment and the internal one containing a
macroscopic number of particles.

    Quantum-mechanically, the position of the pointer's
center of gravity is influenced both by the interaction with
the particles of the pointer itself and by the interaction with
particles outside it. The pointer as a system is described by a
matrix density which is within a short time diagonalized in a
quasiclassical basis. Interference terms, whose presence is
what distinguishes a quantum system from a classical one,
rapidly vanish --- within a time scale which, for certain cases
(e.g., for a system of oscillators), is determined by an
exponential with the number of environmental particles in
its exponent. This is exactly the reason why we do not watch
the interference between the alive state and the dead state of
Schr\"{o}dinger's famous cat. (For a more detailed discussion of
decoherence, see Ref.~\cite{53}.)

    Now, does decoherence solve the measurement problem?
Many believe that it does not. J S Bell, for one, asks what there
is in nature to make a quantum system with a macrobody say:
``I am a system, and you are my environment, and we will
average over you''~\cite{54}. It has also been noted that the
diagonalization of the density matrix without the observation
of the subsystem does not yet imply the emergence of a
mixture of states for the whole system, the point we have
discussed above in connection with the violation of unitarity
in measurement. All decoherence does is to explain why we
human-observers do not see macrobodies interfere.

    Wojciech Zurek~\cite{55} and M B Menskii~\cite{21}, in connection
with the existence of a selected quasiclassical basis of the
eigenfunctions of commuting macroobservables, make an
attempt to relate this choice to evolutionary biological selection.
    In this approach, not only do living beings adapt
themselves to nature, but nature also adapts itself to the living
beings that observe it. How life is viewed from the perspective
of physics was also discussed in Refs~\cite{56, 57}.
    Living beings turn out to be more adapted when they do not ask `dangerous'
questions to the macroworld in their neighborhood, thereby
choosing a special `reference frame in a Hilbert space', so that,
due to decoherence, classical physics with its determinism and
predictability turns out to be approximately valid.

    Other bases also exist, in which Schr\"{o}dinger's cats in the
state of interference walk by themselves next to us. So we see
that quantum physics with all its unpredictability invades the
macroworld. The present author discussed a similar idea in
connection with certain problems in psychology~\cite{58}.
    Decoherence also plays an important role in how we conceive the
macroscopic Universe which, while primordially quantized,
seems to us to be classical with respect to a quasiclassical basis.

    It is important to note that, for macrobodies regarded as
quantum objects consisting of a large number of particles
interacting with a large number of particles in the environment,
decoherence is a consequence of the `entanglement'
between the states of particles in the macrobody and those in
its environment, so that the body itself is described by a
density matrix. This means that a macroapparatus, like any
`classical' body, from the standpoint of quantum physics is
not at all isolated from the surrounding world.

    Furthermore, from a microscopic point of view, an
apparatus macroscopically isolated from the Universe is not
at all isolated from it but, indeed, due to the entanglement of
states, is tied to it more strongly than an individual quantum
particle. This new understanding of classical macroobjects
from the quantum perspective is an important recent
development.

\vspace{11pt}
\section{Everett interpretation}
\label{Sec4}

    Much popularity has recently been gained by the Everett
interpretation~\cite{14}. The idea was disseminated in 1957 and
found initially support from John Wheeler in his paper~\cite{15}
published in the same journal issue.

    Hugh Everett --- a Princeton graduate, theoretical physicist,
and great mathematical talent --- defended his Ph.D.
thesis on a new interpretation of quantum mechanics based
on the assumed objectivity of the wave function. The thesis
was published in 1973~\cite{16}, much later than his famous paper.
After unsuccessful (in his view) discussions with prominent
physicists, he left physics to join a secret Pentagon department,
where his job was the mathematical modeling of
nuclear war strategies; this included, in particular, assessing
the possible consequences of radioactive contamination and
the optimization of nuclear attacks on the USSR.
    He was unsociable and had a reputation, perhaps exaggerated~\cite{59},
for drinking. He died at the age of 51 in 1982. Of his two
children, his son Mark became a rock star, and his daughter
Elizabeth committed suicide by overdosing on sleeping pills in
1996. Her suicide note read that she was ``going to join her
father in another universe''~\cite{60}.

    The Everett interpretation came to be known as the
`many-worlds' or `many-universes' interpretation (which,
incidentally, explains poor Elisabeth's note).

    A few background words first. David Deutsch, one of the
most dedicated Everettians, claimed that the idea of many
worlds described by a single wave function was first
mentioned by Schr\"{o}dinger in one of his lectures, with the
remark that ``the audience will probably consider me crazy'';
anyway, he published nothing on the subject~\cite{61}.

    After his 1957 publication, Everett went to Copenhagen
to discuss his views with Bohr. There, however, his theory
received a rather chilly response. The most incisive was
Rosenfeld, who wrote later to Bell that the assumption of a
universal wave function for the Universe would mean that
``our view of the Universe is that of God''~\cite{62}.

    Nor were such founders of quantum physics as Dirac,
Wigner, and Feynman anything more than lukewarm. Feynman,
in particular, wrote~\cite{62}: ``The concept of a universal
wave function has serious conceptual difficulties. This is so
since this function must contain amplitudes for all possible
worlds depending on all quantum-mechanical possibilities in
the past and, thus, one is forced to believe in the equal reality
[sic!] of an infinity of possible worlds.''

    Norbert Wiener doubted the possibility of introducing the
Lebegue measure in a Hilbert space and advised Everett to
publish his paper as a ``comment on current debates on
quantum theory rather than a definite result''~\cite{62}.

    The situation changed, however, when Bryce DeWitt
derived what came to be known as the Wheeler--DeWitt equation
for the wave function of the Universe. From that
time on, DeWitt was an active defendant and disseminator of
the Everett interpretation. It should be noted that Wheeler
also originally supported the Everett interpretation, in the
hope that it would be of use in developing quantum
gravitation. However, Wheeler denounced this interpretation
in 1977 as ``introducing an infinite number of unobservables
as metaphysical baggage'' and converted to the von
Neumann--Wigner interpretation, in which the consciousness
of the observer is not described by a wave function~\cite{62}.

    Bell's criticism was that the Everett interpretation ``has
many pasts and many futures''~\cite{54}.

    All this notwithstanding, DeWitt's work triggered activity
in `quantum cosmology', in which it is postulated at the outset
that there is a certain Universe wave function which exists
whether or not any observer is present and which has the same
objective status as an electromagnetic field.

    Such cosmologists as Stephen Hawking and Alexander
Vilenkin use this concept in their science and popular science
publications to avoid the problem of the primordial singularity of
the Universe. While Vilenkin is quite unequivocal about
his preferred interpretation (which is Everett's; see Ref.~\cite{19}),
Hawking is somewhat vague and postulates his Universe
wave function as an integral for a certain real (but non-Feynman)
measure over various Euclidean universes.

    It is clear, however, that discarding the Everett interpretation
as false also casts strong doubts on quantum cosmology
studies, unless, of course, we agree with Wheeler that
consciousness is not part of the Universe, and that all the
Universe is defined with respect to this consciousness
according to his concept of the `participating Universe'. But
then, as with any version of the Copenhagen interpretation,
the preparation and measurement of the wave functions and
the observer-induced wave packet collapse --- that is, the
concepts ignored by the quantum cosmologists --- should be invoked.

    Advocates of the Everett interpretation include D Deutsch
(UK)~\cite{63}, B Carter (Australia)~\cite{64}, M Tegmark (US)~\cite{65},
Don Page (Canada)~\cite{66}, M B Menskii (Russia)~\cite{20, 21}, and
Lev Vaidman (Israel)~\cite{67}.

    In 2007, a number of events commemorated the 50th
anniversary of Everett's seminal paper, including a special
Nature issue~\cite{68}, a popular article in the Scientific
American~\cite{69}, two international conferences, and a BBC film on
parallel universes. Note also that the Everett parallel
universes are being promoted in science literature by
supporters of string theory, `the theory of EVERYTHING'
(see, for example, Ref.~\cite{70}).

    We will now proceed to directly discuss the Everett
interpretation. In doing so, we will quote and refer to both
Everett himself and the current Everettians. From its very
start, the Everett interpretation positioned itself as opposing
the Copenhagen interpretation.

    (1) A wave function possesses the same objective reality --- and
is an observer-independent --- as an electromagnetic field.

    (2) There is a universal wave function, i.e., the wave
function of the Universe. In addition, Tegmark, an Everettian,
says that individual particles have `epistemological'
wave functions, which are what experimentalists deal with;
however, these functions also exist objectively~\cite{65}.

    (3) The wave function satisfies the Schr\"{o}dinger equation
and evolves unitarily. There is no need to postulate any wave
packet reduction --- i.e., a nonunitary operation --- because all
the observed properties follow from the Schr\"{o}dinger equation.

    (4) A measurement of a physical quantity in the case of
the superposition of wave functions `splits' the initial states of
both the observer and the particle into a multitude of states
belonging to different and noninteracting `worlds' or universes.
    In addition to the splitting, there is a possibility that,
due to the time reversal of the Schr\"{o}dinger equation, the
`worlds' will merge, which distinguishes this theory from
standard quantum physics, in which measurement violates
time reversal. Therefore, the statement that the Everett
interpretation leads to the same consequences as the Copenhagen
interpretation is, of course, wrong.
    The only question is how to check this experimentally.

    (5) Quantum mechanics applies to both microscopic and
macroscopic phenomena. Classical physics provides an
approximate description of those quantum objects with a
large number of particles that satisfy the decoherence property.

    Actually, it is clear why Schr\"{o}dinger could give his full
support to the above points. There indeed are arguments
which Schr\"{o}dinger could make against those criticizing his
view that waves exist objectively and the electron constitutes a
wave. His critics argued that the electron is always observed
as a pointlike particle and that it is different from the
electromagnetic wave, for which different radio receivers
detect something at any point of its front. An electron is not
observed as smeared over any front surface and, hence, the
electron wave is a `probability wave' for finding the electron
at any place in space.

    Schr\"{o}dinger could object, following Everett, that an
electron, while remaining pointlike particle, resides simultaneously
(and, of course, without any self-multiplication) in
different places of space in different worlds --- the worlds in
which the states of the observer are also different and
correspond to different observational results, and which are
independent for different states of the observer and not
interacting between themselves.

    A few words on Everett's paper~\cite{14} are in order. All the
mathematical formulas it contains are known in ordinary
quantum mechanics as well. Where the difference lies is in
how they are interpreted. Of key importance is the concept of
the `relative wave function' of a subsystem of a quantum
system. If a complex system S consists of two subsystems, $\rm S_1$
and $\rm S_2$, the Hilbert space for S is the tensor product of the
Hilbert spaces of the subsystems. If these spaces contain
complete orthonormalized sets
$\{ \xi_{i}^{\rm S_1} \} $, $\{ \eta_{j}^{\rm S_2 } \} $,
the total state of the system~S can be represented as
    \begin{equation} \label{EQ11}
\Psi ^{\rm S} =\sum _{i,j} a_{ij}\, \xi _{i}^{\rm S_1 }  \eta _{j}^{\rm S_2 } .
\end{equation}
    From this it follows that, although S occupies a definite state
$\Psi_{\rm S}$, the subsystems are not in definite states with respect to
each other. However, for a certain choice of state for one
subsystem, we can determine the corresponding relative state
of the other. For example, choosing $\xi_k$ as a state of $\rm S_1$,
the relative state in $\rm S_2$ has the form
    \begin{equation} \label{EQ12}
\Psi \left( {\rm S_2 } ;\; {\rm rel}( \xi_{k} , {\rm S_1 } ) \right) =
N_k \sum \limits_j a_{kj} \, \eta_j^{\rm S_2} ,
\end{equation}
    with $N_k$ being the normalization factor. Paper~\cite{14}, therefore,
goes on to argue: ``It is meaningless to talk about the absolute
state of a subsystem -- one can only talk about the state
relative to a given state of the another subsystem.''

    Comparing this description with the situation in the
theory of relativity, where the length of a body is defined
relative to a reference frame and is otherwise indefinite, we
note --- by considering the system particle--apparatus--observer,
in which, according to Everett, the observer is also
described by a wave function --- that the observer's wave
function is defined relative to the apparatus wave function,
which, in turn, is defined relative to the wave function of the
particle. It should be noted, though, that prior to measurement,
the observer --- according to equation~(10) of Everett's paper ---
was characterized by a certain `absolute' wave
function determined by the memory of events, so that the
system involving the observer plus the quantum object before
their interaction during the measurement is, according to the
measurement Hamiltonian, described by (Eqn~(10) in Ref.~\cite{14})
    \begin{equation} \label{EQ13}
\Psi^{\rm S+O} = \Phi _{i}\, \Psi^{\rm O} [\ldots]\,.
\end{equation}

    We emphasize the following properties concerning the
measurement of a quantity described by the operator of the
observable $ \hat{A} $ with eigenfunctions $ \Phi _{i} $.

    If the quantum system was found itself in an eigenstate of
the operator $ \hat{A} $, then, after the measurement, the total wave
function transforms into the function (Eqn~(11) in Ref.~\cite{14})
    \begin{equation} \label{EQ14}
\Psi^{\rm S+O} = \Phi _{i}\, \Psi^{\rm O} [\ldots, \alpha_i]\,.
\end{equation}
    However, if the system resides initially in a superposition of eigenstates,
$\sum _{i} a_{i} \Phi_{i} $, the final state after the measurement is
given by (Eqn (12) in Ref. [14])
    \begin{equation} \label{EQ15}
\Psi^{\rm S+O} = \sum _{i}\, a_{i} \Phi_{i} \Psi^{\rm O} [\ldots, \alpha _{i} ] ,
\end{equation}
    where the result $ \alpha_i $ appears in the memory of the observer as an
apparatus reading and is associated with that eigenvalue of
the operator $ \hat{A} $ which corresponds to
the eigenfunction $ \Phi_{i} $ of this operator.

    And here we come to the important point in interpreting
this obvious property of measurement.
    Everett insists on the following.

    (1) Quantum physics describes all happening in microworld and
macroworld alike. There is no need to follow Bohr
in dividing the world into classical and quantum domains and
describing the macroapparatus and the observer classically
and the microobject quantum mechanically.

    (2) Consciousness is totally irrelevant; the observer can be
replaced by an automation or by a computer with memory.

    (3) All the terms in superposition~(\ref{EQ12}) are equally real.
The wave function collapse concept is altogether unnecessary.
In particular, it cannot be argued, following Copenhagen
style, that, of all the sum, only one term is realized with a
probability determined by the Born formula as a result of the
measurement. The initial wave function undergoes a `splitting' into
superposition of states~(\ref{EQ12}), with each term
describing a real `world' (or branch) in which the observer
views a definite, world-specific measurement result.

    (4) The worlds do not interact with each other, so an
observer in one world knows nothing of another world.

    It is shown further that, when measuring the same
quantity again, the observer in a fixed world will, due to the
unitary evolution in each branch, survey the same result.

    Next, the concept of a measure for the superposition
$ \sum_{i} a_{i} \Phi_{i} = \alpha \; \Phi^{'} $, namely
    \begin{equation} \label{EQ16}
m (\alpha ) = \sum \limits_{i=1}^{n} m(\alpha_{i} )
\end{equation}
is introduced and it is proved that the countable additivity of
this measure fixes it as
    \begin{equation} \label{EQ17}
m (\alpha_{i} )=c\, a_{i}^{*} a_{i} ,
\end{equation}
    where $c$ is a constant --- that is, we arrive at the Born
probability formula of quantum physics.

    Finally, an analysis is made of a system of many quantum
objects, $\rm S_1, S_2, \ldots , S_n$, identical in the sense of being all
described by the same wave function
$ \Psi^{\rm S_1} = \Psi^{\rm S_2} = \ldots \Psi^{\rm S_n} =$
$ \sum_i a_i \, \Phi_i$.
    Before measurement, the wave function of the system
and the observer with memory is written out as
    \begin{equation} \label{EQ18}
\Psi^{\rm S_1 + S_2 + \ldots + S_n + O} =
\Psi^{\rm S_1} \Psi^{{\rm S}_2} \ldots \Psi^{{\rm S}_n} \Psi^{\rm O} [...]\, .
\end{equation}
    After $r \ (r \le n)$ measurements, assuming that the measurements of
one and the same quantity~$A$ are performed in the order
$ {\rm S}_1, {\rm S}_2, \ldots , {\rm S}_r$, we obtain for n systems
the following expression
    \begin{equation} \label{EQ19}
\Psi_{r} =\sum \limits_{i,j,...,k} a_{i} a_{j} ... a_{k} \Phi_i^{{\rm S}_1}
\Phi_j^{{\rm S}_2} \ldots \Phi_k^{{\rm S}_r}
\Psi^{{\rm S}_{r+1} } \ldots \Psi^{{\rm S}_n } \Psi^{{\rm O} }
[ \alpha_{i}^{1} \alpha_{j}^{2} \ldots \alpha_{k}^{r} ]\,.
\end{equation}
    The function $ \Psi_{r} $ is interpreted as a superposition of states:
    \begin{equation} \label{EQ20}
\Psi^{'}_{i j \ldots k} = \Phi_i^{{\rm S}_1}
\Phi_j^{{\rm S}_2} \ldots \Phi_k^{{\rm S}_r}
\Psi^{{\rm S}_{r+1} } \ldots \Psi^{{\rm S}_n } \Psi^{{\rm O} }
[ \alpha_{i}^{1} \alpha_{j}^{2} \ldots \alpha_{k}^{r} ]\,,
\end{equation}
    each of which describes the observer with a certain sequence
in memory:
    \begin{equation} \label{EQ21}
\left[ \alpha_{i}^{1} \alpha_{j}^{2} \ldots \alpha_{k}^{r} \right] .
\end{equation}
    Note that the state of the system with respect to the memory
of the observer is a product of the system's eigenfunctions
$ \Phi_i^{{\rm S}_1}, \Phi_j^{{\rm S}_2}, \ldots, \Phi_k^{{\rm S}_r} $,
whereas the remaining wave functions of
the system are left the same.

    In the memory of the observer, a configuration like~(\ref{EQ21})
exists, which is random for an observer who fixed a certain
world and nonrandom in nature. Some numbers in the
sequence are the same, and some are not. Different branches
will possess different sequences.

    From the above, it is concluded that the observer (who
can also be an automation) shows in one of the worlds exactly
what is actually shown by the apparatus measuring the
quantum object. And this without any use of the wave packet
collapse concept!

\vspace{7pt}
\subsection{Discussion topics}
\label{SubSec41}

    Unfortunately, both the proponents and opponents of the
Everett interpretation are as often as not accurate or even
incorrect when interpreting his world splitting idea. It is said
about the splitting of the particle and the observer into
`copies' or `twins' as of something tantamount to their
`multiplication' at measurement. This view is inconsistent
with the laws of conservation of energy and charge and is
physically meaningless.

    Multiplication occurs in Hilbert space; what multiplies is
the states of an object, not the objects themselves.

    An electron that is single before remains so after
measurement (as does the observer: no twins there!) and is
described at once by many relative states. The observer views
one result in one of these states, and another in the other state.
This means that a man in one and the same (nonmultiplying)
body resides in many states at once, so that he surveys one
result in one state, another in another, yet another in yet
another, and so on. It is not without reason that DeWitt
himself called this situation schizophrenic~\cite{16}.
This is something to consult a psychiatrist on, though, who may have
something useful to say to a physicist on this.

    Pictures by Escher or some works painted by Salvador Dali may also
serve to illustrate the splitting idea.
    In Escher's {\it Wild Geese}, for example, the birds look white when
they fly in one direction and black when they fly in the opposite
direction. So, one and the same picture can be called {\it White--Black Geese}
since there are two states --- and without the picture being multiplied.

    By and large, the Everett interpretation cannot be blamed
for violating conservation laws. It is the problems of preferred
basis and probability that are the subject of criticism.

    {\it The problem of preferred basis.}
    In the Copenhagen interpretation, if the observer chooses to measure an
observable described by operator $ \hat{A} $, he puts the apparatus
readings into correspondence with the eigenfunctions and
eigenvalues of this operator. An Everettian observer makes
no choices: all that happens does so according to the
Schr\"{o}dinger equation. But then the wave function of the
particle can be expanded on any orthonormal basis, not only
in terms of the one chosen in the Copenhagen interpretation.

    The answer of an Everettian uses the concept of
decoherence. If the apparatus consists of a large number of
particles, then a preferred basis exists which allows one,
owing to decoherence, to speak of the approximate absence
of interference between various terms in the superposition of
the states of the macroobject.

    It then follows that the split `worlds' do not interact with
each other, and the observer residing in one state knows
nothing of the existence of other states. It is also noted that,
due to the interaction of the body of the observer with other
macroobjects, the states of the Universe ultimately split,
making it necessary to speak of `many universes'~\cite{65}.
    Here, however, the same inaccuracy arises: this is not a multitude of
universes but rather one Universe occupying a multitude of
states. Recalling Schr\"{o}dinger's cat, it can be said that he is
alive in one universe and dead in another, etc.

    By way of objection, though, some measurements are
performed with nonmacroscopic apparatuses. An example is
the Stern--Gerlach electron spin experiment, in which silver
ions that play the role of an apparatus move along
quasiclassical trajectories in one direction or another, and
decoherence hardly plays any role.

    {\it The problem of probability.}
    In the Everett interpretation,
all events occur deterministically, and there is no random
quantity which takes one or another values with some
probabilities. On the other hand, as already noted above,
Everett obtained a certain countably additive measure on
superposition elements, which is equal to the Born rule value.
What is the meaning of this measure?

    One attempted answer is that an individual observer in
one of the worlds, who knows nothing of the existence of
other worlds, perceives herself as having appeared by accident
in this world. Many of the Everettians, though --- unlike
Everett himself --- find it necessary to resort to consciousness,
which is, of course, some capitulation to Copenhagen\,... .

    The following example, proposed by Tegmark~\cite{65},
illustrates things nicely.

    A surgery patient wakes up from general anesthesia and
finds herself in a ward with a definite number. If we consider
general anesthesia as a quantum measurement, there are
many copies (states) the patient can be in, of which she
conceives only one. After another operation under general
anesthesia, the patient may find herself in a ward with a
different number. So the number is a random quantity for the
patient. But what is the probability of identifying the observer
with a fixed copy? If all the worlds are equally existent, we
could assign the same probability to each of the worlds
conceived by the observer, which is at odds with quantum
physics. Thus, if a state is an eigenfunction of the operator of
the $z$-component of electron spin, and the quantity being
measured is the projection of the spin operator onto a certain
axis other than normal to the $z$-axis, then the probabilities of
this outcome or others are determined by the squares of the
cosine and sine of the angle between above-specified axes, and
these in no way equal $1/2$. If one tries to express probability in
terms of the outcome frequency of a certain `world', this
frequency can be arbitrary, a point with which the Everettians
themselves agree. Thus, they denounce von Mises's frequency
(or statistical) interpretation of probability.

    An alternative proposal put forward by Deutsch~\cite{71} is a
subjective, frequency-unrelated probability interpretation
used in decision making theory. It is this probability which
is identified with that obtained from the Born formula.

    By way of example, if you are told that, depending on your
decision, you can win a thousand dollars, five hundred
dollars, or nothing in three cases, then, unaware of the
frequency of wins, you rely on your subjective assessments
to make a decision. Of course --- noting that you know
nothing of any frequencies! --- this requires considering a
certain objective property you `know' of, in order to avoid
frustration. This can be, for example, the symmetry or
perhaps some distortion in the symmetry of the coin (in the
heads or tails game), or this can be the past experience of the
decision-maker.

    The Everett interpretation involves a multitude of worlds
with an observer in them. However, the observer perceives
`herself' (or `me') as `embodied' in one of the worlds, whereas
his other states (or `copies', to use an Everettian term) observe
(or conceive of) nothing. What is it that determines this
`embodiment' or this `preference' of one world over another?

    The only thing that features in the mathematical theory is
the coefficients in the superposition or weights as their
absolute values squared. This makes it tempting to consider
these different coefficients as something like elements of the
payment matrix in a quantum game. The larger the coefficient,
the more preferred this world for the observer.

    On this, Tegmark~\cite{65} says that the observer undergoing
splitting will consider the Born formula as determining a
certain measure on the worlds. This measure is determined
based on the properties of Hilbert space. If there is initial
wave function before and a different one after the measurement,
nothing other than a scalar product of these two wave
functions and its absolute value squared (which is similar to
the absolute value squared of the coefficient in the superposition)
can be proposed as such a measure. The weights will
be different for various wave functions characterizing worlds.
Is it possible, however, to agree with Saunders~\cite{72, 73} that, in
Everett's theory, which insists on unitary evolution, the
probability amplitudes of worlds are --- like weights --- unobservable,
unlike the probability amplitude in the Copenhagen
interpretation? Everett himself, judging from his paper~\cite{14},
did not agree with this. According to formulas (\ref{EQ19})--(\ref{EQ21})
of the present paper, the `memory' of the observer stores the
observation `frequencies' of those or the other worlds. Were
this otherwise, it would be completely impossible to speak of
quantum mechanics and its verification experiments which
involve, in particular, the diffraction and interference of
electrons (when the frequency with which a particle hits one
point or another is measured).

    However, does this mean that, to use Everettian language,
we survey the fingerprints of different worlds in a picture of a
diffraction pattern? And to which world (or universe) does
this picture now belong? Can it be said that a multitude of
past worlds meet in the present one?

    Furthermore, as shown by Graham~\cite{25}, the frequencies
that are determined, as suggested by Everett, from the
number of coincident values in the memory of the apparatus
[see Eqns~(\ref{EQ19})--(\ref{EQ21})] may be arbitrary, which is precisely
the reason for Saunders's statement.

    Returning to Deutsch's idea on decision making theory,
we note, however, that, while in decision theory as applied to
business there is nothing unclear, it is impossible to understand
why the observer prefers to be `embodied' in one world
and not in another, given that he will have neither benefit no
matter whether the electron spin is up or down.

    Another possible line of reasoning is possible here, which
is similar to the one we used when discussing the Copenhagen
interpretation.

    According to the Bohr--Fock principle of relativity with
respect to the means of measurement, augmented by Everett's
notion of a relative state of a subsystem, a definite reading of
the apparatus signifies a `relation' between the particle and
the observer. Adopting the `neo-Everettian' viewpoint that
the observer's consciousness is important in identifying his
state with one of the split states, whereas other states exist but
are not perceived by him, we could say the following. An
electron, which, while possessing something like freedom of
choice, is unable to make decisions because of the absence of
consciousness, is characterized by many `relations' to the
observer. The Born measure can be understood as defined on
these relations (characterizing, by definition, both the
electron and the observer). As an electron with the same
wave function is repeatedly the subject of measurements, the
observer acquires a `tendency' to find himself more often in a
world with a larger Born weight. The reason is that repeated
measurements of a single electron with a preassigned wave
function have the consequence that, as noted above, the
observer perceives himself in the same world. If the weight
does not differ significantly from unity, the observer is more
likely to see herself in the same place (the same world).

    Tendency is a fragile thing, however, and can be
denounced, and then other, even small-value, weights are
realized, defined on the relations between the observer and
the electron. It is the fragile nature of the tendency to identify
the state of the observer with one of the worlds, which is the
reason for quantum-mechanical indeterminism. But is the
choice of a world something dependent on the will of the
observer, as the advocates of the decision making approach
claim, or is it not? Because the Born measure describes not
only the observer but also the subject (observer)-object
(particle) unity, the answer is likely to be negative.

    It is not (not only!) by her will that the observer conceives
herself in a fixed world. Consciousness as a choice of one of
the alternatives (here we agree with Menskii~\cite{20}) or worlds
should not be treated as a property of the brain or of
something `privatized' by a given observer but as a nonlocal
property that embraces both the observer and the object of
cognition. As the philosophy of intuitivism formulates it,
``the object of knowledge is immanent in knowledge''~\cite{74}.

    Cognition is an objective process which, like time,
embraces both the observer and the observable. Consciousness is
a nonlocal property of the electron and the observer
and is not related to the brain alone. The Born formula
exhibits a characteristic of this nonlocal property.

    Here, we should agree with DeWitt~\cite{16} that quantum
probability is a special kind of probability, which does not
reduce to any classical probability, in particular, to the
subjective probability involved in decision making theory.
The equal existence of worlds does not mean that, similar to
throwing a die with six identical faces, the observer can adopt
the classical definition of probability and assign the same
probability to all worlds. Is this answer convincing? Hardly.
All this reasoning about an observer, decisions making by
her, and consciousness etc. have no relation to the Schr\"{o}dinger
equation and quantum-mechanical mathematics (thought
of at the very start as all-describing) and do nothing more
than revive the old wave packet collapse debate --- but in a
much more vague and nebulous manner this time.

    The role of consciousness in the Everett interpretation is
also discussed by Brandon Carter~\cite{64}, who rejects Deutsch's
decision making approach.

    The dilemma covering the equal existence of different
worlds in a superposition and the different probabilities of
their observation according to Born's measure is proposed to
be resolved by analogy with the Russian roulette game, an
invention of Russian white-guard officers, in which a revolver
is loaded with (say) four blanks and one bullet. From the
point of view of outcomes, the probability of some outcome is
one in five, but from the point of view of a potential suicide,
the probability of remaining alive is four in five, and of dying,
one in five.

    Of importance in this context is the identification and
distinguishing of worlds. The observer identifying herself
with one of the copies can identify some worlds and
introduce different weights for the group of worlds she
identifies and for those she distinguishes. It remains unclear,
though, how to obtain exactly the Born formula, and why, in
fact, this should be the Born formula, and why it should be
obtained at all, for that matter. The rules of quantum physics
are unlikely to be derived from any classical analogs
whatsoever.

    Carter also levels criticism at the concept of `splitting
worlds' by arguing that worlds `exist eternally' and that all
there is to talk about is distinguishing and identifying them.
Clearly, the Schr\"{o}dinger equation should be applied to all (an
infinite number of) wave functions, which is reminiscent of
Feynman's critical remark that the Everett theory loses much
of its validity because of the necessity to consider an infinite
number of probability amplitudes. Criticisms of, and indeed
accusations of inconsistency against, the subjective prob-
ability approach come from Kent~\cite{75}. In any case, the
present author's view is that this approach, while interesting,
is not sufficiently grounded to enter quantum physics textbooks.

    Thus, we should agree with Kent~\cite{75} that Everettians are
often contradictory in their publications and have as yet no
consensus on how the Everett interpretation should be interpreted.

    Some authors, for example DeWitt, emphasize that
consciousness does not play any role, while others argue the
opposite and try to employ decision making theory. Finally,
Deutsch~\cite{63} himself, the chief pro-Everettian, notes, quite
rightly, that such aspects as the smearing out of the wave
packet, a situation where worlds obviously interact and where
interference is involved, have not been given sufficient
analysis. Even the highly popular two-slit experiment with a
quantum particle has not been treated this way.

    Opinions differ not only on the `splitting' of worlds but
also on their `merging' (see above). Some agree that merging
is necessary~\cite{65} and some do not, because they do not
consider the merger of macroscopic decoherent worlds possible.

\vspace{11pt}
\section{Conclusion}
\label{Concl}

    To summarize, our preference goes, of course, to the
Copenhagen interpretation, because it is worked out more
and agrees with all observations in the microworld.
    Materialistic prejudices that determine the rejection and prevent
adoption of the Copenhagen interpretation by some physicists can hardly
be considered as a serious argument~\cite{76}.
Unfortunately, strengthening the case of materialism in the
eyes of many scientists is their desire to distance themselves from various
kinds of parascience that discredits nonmaterialistic philosophy.

    It is, however, curious --- and bitterly ironic --- that
anti-Copenhagenists themselves, with their self-proclaimed `realistic'
many-worlds interpretation, find themselves obliged to
speak of consciousness and of the necessity to identify the
observer with something fixed in a certain world, thereby
demonstrating, in a sense, the convergence of above interpretation.
    We put aside, of course, the wild talk about Everettian parallel-world
mysticism typical of fantasy fiction, cinema, and mass media~\cite{69}.

    But also in the Copenhagen interpretation itself, if it is
also considered realistic, with observer-independent quantum
reality described by operators and not by numbers (as explained by Aage Bohr),
time evolution should, as mentioned above, be described in the Heisenberg
representation.
    Thus, quantum reality in the language of operators contains
--- due to the possibility of spectral decomposition in projectors into
the eigenstates --- not a single world but many worlds at once, and evolution
is the evolution of all these worlds simultaneously.
    This is, of course, very close to Everett's picture.

    Still, the Copenhagen interpretation, with its emphasis on
the role of consciousness, may have something new to say
about the relation between consciousness and the brain. This,
according to von Neumann, requires drawing a boundary
between the two. The thirty plus years of discussions
conducted by the present author at the Institute of Human
Brain of the Russian Academy of Sciences (St. Petersburg),
while not producing any definite results, have established a
number of points of agreement.

    (1) Brain specialists have become interested in the nonlocality,
or entanglement, of quantum states of many particles --- properties
that have analogs in the nonlocal aspects of the operation of
the brain as a whole.

    (2) The main question common to philosophy, psychology, and brain
theory is formulated as a psychophysical mental issue.
    How does it happen that obviously nonmaterial mental processes control
and affect the physical and chemical processes in the brain and the body?
    A possible answer~\cite{77} can be hidden in quantum physics with its
negative experiments in which, while negative information
involves no energy or momentum transfer from the observer
(apparatus) to the quantum object being observed, it nevertheless changes
the energy and momentum of the object (to a degree allowed by the uncertainty
relation).
    Thus, nonmaterial consciousness, which does not possess physical energy
and momentum, can lead, due to its main property (consisting in
`cognition of information'), to physical and chemical consequences.

    (3) What is unique about my conceiving of processes in
my body is that consciousness considered as a quantum
apparatus is not material --- like the mental in general ---
and, hence, does not consist of quantum particles, and,
further, cannot be thought of in terms of decoherence.

    Information on the physical characteristics of the body is
formulated not in physical terms like the position of the
pointer in a measuring instrument but in purely mental terms
like joy, pain, and feelings like these. That this `apparatus'
possesses no ability to decoherence means that, unlike an
ordinary apparatus used in physical experiment, it can also
observe other than classical observables\,...
    I can observe Schr\"{o}dinger cats in my body.
    Curiously, any external observation of the brain using physical
apparatuses in some way or another prevents me, due to the decoherence that
operates in this case, from surveying this interaction between
my consciousness and my body. It is well known, though, that
you can cut the brain in whatever way but you will never watch consciousness.
    `Quantum conspiracy' seems to be an appropriate term here.

    A major difficulty in discussions with biologists and
medical people is their traditional materialistic position
determinate mechanistically in the spirit of Newton, a
position which is less characteristic of physicists. There have
been two definite exceptions, though: N P Bekhtereva, a
person of great freedom of thought, always interested in the
philosophical assessment of the results of her long-running
brain studies, and S V Medvedev, Director of the Institute of
Human Brain and a physics graduate.

    To conclude, our view is that, while the many-world
interpretation is currently not sufficiently elaborated and is
rather a compilation of less than convincing and sometimes
poorly connected ideas, the current understanding that most
of the Universe is occupied by unobservable (invisible) dark
matter and dark energy calls for more serious attention to a
theory in which the existence of many unobserved worlds is
assumed as an axiom. The question is how these worlds
gravitate.

\vspace{7pt}
\noindent
{\bf Acknowledgments}\\
This work was carried out in cooperation with the Copernicus
Center for Interdisciplinary Studies in Krak\'{o}w, Poland,
under financial support from the Templeton Foundation.


\end{document}